\documentclass[twocolumn,prl,superscriptaddress]{revtex4-2} 

\usepackage[T1]{fontenc}

\usepackage{amsmath,amsfonts,bm,amssymb}
\usepackage{graphicx}  
\usepackage{color}
\usepackage{extarrows}
\usepackage{enumitem}
\usepackage{empheq}
\usepackage{tensor}

\newcommand{\ud}{{\mathrm d}}

\newcommand{\B}{\mbox{\tiny B}}

\newcommand{\tB}{\mbox{\tiny B}}

\newcommand{\tS}{\mbox{\tiny S}}

\newcommand{\T}{\mbox{\tiny T}}

\newcommand{\SB}{\mbox{\tiny SB}}

\newcommand{\la}{\langle}
\newcommand{\ra}{\rangle}

\newcommand{\nl}{\nonumber \\}
\newcommand{\be}{\begin{equation}}
\newcommand{\ee}{\end{equation}}
\newcommand{\bsube}{\begin{subequations}}
\newcommand{\esube}{\end{subequations}}
\newcommand{\Eq}[1]{Eq.\,(\ref{#1})}
\newcommand{\Eqs}[1]{Eqs.\,(\ref{#1})}

\newcommand{\RN}[1]{%
  \textup{\uppercase\expandafter{\romannumeral#1}}%
}

\allowdisplaybreaks[1]

\definecolor{darkblue}{RGB}{0, 56, 102}

\usepackage{hyperref}  
\hypersetup{hidelinks,
	colorlinks=true,
	allcolors=black,
	pdfstartview=Fit,
	breaklinks=true
}

\begin{document}

\title{Nonperturbative Open Quantum Dynamics Bypass Influence Functional
}

\author{Yu Su}
\email{suyupilemao@mail.ustc.edu.cn}
\author{Yao Wang}
\affiliation{
  Hefei National Research Center for Physical Sciences at the Microscale, University of Science and Technology of China, Hefei, Anhui 230026, China
}
\author{Wenjie Dou}
\affiliation{
  Department of Physics and Department of Chemistry, School of Sciences, Westlake University, Hangzhou, Zhejiang 310024, China
}

\date{\today}

\begin{abstract}
  An ordered moment approach to exact open quantum dynamics is presented, which bypasses the Feynman--Vernon influence functional formalism. The hierarchical equations of motion are constructed using Wick's contraction, which follows specific orderings of the bath's creation and annihilation operators. Our approach moves beyond the traditional influence functional formalism, offering a more intuitive and direct framework, and extends the applicability of theory to nonlinear system--bath coupling scenarios.
\end{abstract}

\maketitle

Quantum mechanics of open systems has been a focus of research since the early days of quantum theory and has become an essential component of modern science, particularly in chemical physics, condensed matter physics, and quantum information science \cite{Wei21}. The core problem of open quantum dynamics is to determine the time evolution of a system ($H_{\tS}$) coupled to one or more environments ($H_{\B}$), where quantum entanglement, dissipation, and transport arise due to the system--environment interaction ($H_{\SB}$). The system and environment together form a closed system with time evolution governed by the Schrödinger equation, or the equivalent von Neumann--Liouville equation ($\hbar \equiv 1$),
\begin{align}\label{Liouville}
  \dot\rho_{\T}(t) = -i [H_{\T},\rho_{\T}(t)] = -i [H_{\tS} + H_{\B} + H_{\tS\tB},\rho_{\T}(t)].
\end{align}
Here the system--bath interaction is generally expressed as direct product of system and bath interacting modes, namely $H_{\SB} = \sum_{a}\hat Q_a\otimes\hat F_a$. The most prevalent model of environment is the Gauss--Wick bath, denoted as \cite{Fey63118,Cal83374,Leg871} 
\begin{align}\label{linear}
  H_{\B} = \sum_{j=1}^N\frac{\omega_j}{2}(\hat p_j^2 + \hat x_j^2) \quad\text{and}\quad \hat F_a = \sum_{j=1}^N c_{aj}\hat x_j.
\end{align}
Feynman and Vernon introduce the influence functional formalism to establish a universal and nonperturbative framework for open systems coupled to Gauss--Wick baths \cite{Fey63118,Cal83374}. They show that the influence functional describing all the non-Markovian effects of the environment on the system only depends on the bath correlation functions, $\la\hat F_a^{\B}(t)\hat F_b^{\B}(0)\ra_{\B}$, where $\hat F_a^{\B}(t)$ is the Heisenberg operator defined with $H_{\B}$ and $\la(\cdot)\ra_{\B}$ is the average over the thermal state of the bath. Various numerical methods are proposed to solve the influence functional. There are two main widely used approaches. One is evaluating the influence functional in the path integral representation directly, such as quasiadiabatic path-integral method. The other is the differential equivalence of the influence functional, the hierarchical equations of motion (HEOM) method \cite{Tan89101}, which is based on the exponential decomposition of the performing derivative on the influence functional. With the development of efficient algorithms, the influence functional formalism and related methods has been applied to a wide range of problems, making great progress in the study of condensed matter physics, chemical physics, and quantum biology. 

However, the influence functional formalism has some theoretical limitations. On the one hand, the formalism focuses on the influence of bath on system's dynamics, which is not direct to acquire bath's dynamical information. For example, in order to obtain the system--bath correlated quantities, such as heat current and bath absorption spectra, one has to employ the nonequilibrium Green's function formalism or the system--bath entanglement theorem. On the other hand, the influence functional is only analytically solvable for the linear system--bath coupling since higher order couplings, for example $\hat Q_{ab}\hat F_a\hat F_b$, will lead to algebraic complexity when performing Wick's theorem \cite{Cas911960}. Such nonlinear couplings play an important role in modeling physical mechanism, such as the superconduct phenomena in the generalized Holstein mode \cite{Rag23L121109,Han24226001}, the Duschinsky rotation in the vibronic spectroscopy, and so on.

Here, we present a general ordered moment approach to exact open quantum dynamics, which overcomes the limitations of influence functional and its equivalent HEOM method. 
We introduce the key quantity, the ordered density operators, as the dynamic variables and construct the following time evolution. We will show later that the equations we derived are identitical to the HEOM in the linear coupling case. Furthermore, our approach can be easily extended to the nonlinear system--bath coupling scenarios, which is beyond the capability of the influence functional and HEOM. This Letter is organized as follows. We firstly outline and exemplify our theory a simple case, where the bath is of the Gauss--Wick type and has discrete modes, i.e., \Eq{linear} with $N$ being finite. Nextly, we generalize our theory to the continuous bath modes, where quantum dissipations emerge. We then discuss the generalizations of nonlinear system--bath interactions---beyond the influence functional formalism. Finally, we summarize our work and discuss the future directions.

We start with the total system-plus-environment Hamiltonian with a single interaction mode, $H_{\T} = H_{\tS} + H_{\B} + \hat Q\hat F$, where the bath Hamiltonian and interaction mode are given by $H_{\B} = \sum_{j=1}^N\frac{\omega_j}{2}(\hat p_j^2+\hat x_j^2) = \sum_j\omega_j\hat a_j^\dagger\hat a_j$ and $\hat F = \sum_{j=1}^Nc_j\hat x_j = \sum_j c_j(\hat a_j + \hat a_j^\dagger)/\sqrt{2}$, respectively. 
Here, $\hat a_j^\dagger$/$\hat a_j$ is the creation/annihilation operator defined from the harmonic oscillator. Note that generalizing our discussions to multi-mode case [\Eq{linear}] is straightforward. The system--bath hybridizing process adopts the initial state being the direct product of an arbitrary system state and the bath canonical state, namely
\begin{align}
  \rho_{\T}(0) = \rho_{\tS}(0)\otimes \rho_{\B}^{\rm eq}(\beta) \equiv \rho_{\tS}(0)\otimes e^{-\beta H_{\B}}/Z_{\B}.
\end{align}
Here, $\beta$ is the inverse temperature and $Z_{\B} = {\rm tr}_{\B}e^{-\beta H_{\B}}$ is the bath canonical partition function. 

To proceed, we introduce the thermofield decomposition to map the original canonical thermal state into an effective vacuum. Specifically, we expand \cite{Ume95}
\begin{align}\label{a2b}
  \hat a_j \equiv \sqrt{\bar{\mathfrak{n}}_j+1}\hat b_j + \sqrt{\bar{\mathfrak{n}}_j}\hat b'_j,
\end{align}
such that the effective vacuum state $|{\rm Vac}\ra$ satisfies 
\begin{align}\label{vac}
  \hat b_j|{\rm Vac}\ra = \hat b'_j|{\rm Vac}\ra = 0
\end{align}
and the Heisenberg evolutions of $\hat b_j$ and $\hat b_j'$ are given by
\begin{align}\label{bt}
  \hat b_j(t) = \hat b_j(0)e^{-i\omega_jt},\quad \hat b_j'(t) = \hat b_j'(0)e^{i\omega_jt}.
\end{align}
In \Eq{a2b}, $\bar{\mathfrak n}_j \equiv 1/(e^{\beta\omega_j}-1)$ is the average occupation number of the $j$-th bath mode. The vacuum is defined in the doubled space of the original bath Hilbert space, namely \cite{Ume95}
\begin{align}
  |\mathrm{Vac}\ra \equiv \frac{1}{\sqrt{Z_{\B}}}\bigotimes_j\sum_{n_j=0}^\infty{e^{-\beta\omega_jn_j/2}}|n_j\ra\otimes|n_j\ra'.
\end{align}
Here, the states $\{|n_j\ra\}$ are the eigentates of the original bath $H_{\B}$, whereas $\{|n_j\ra'\}$ are that of the auxiliary bath $H_{\B}' = -\sum_j\omega_j\hat a_j'^\dagger\hat a_j'$, with $\hat a_j' = \sqrt{\bar{\mathfrak{n}}_j+1}\hat b_j' - \sqrt{\bar{\mathfrak{n}}_j}\hat b_j$. Note that the auxiliary bath $H_{\B}'$ actually is the time--reversed bath of $H_{\B}$, with basic eigen-frequecies being $\{-\omega_j\}$ \cite{Ume95}. Tracing over the auxiliary bath results in the original bath canonical state. 

As a result, the bath Hamiltonian becomes
\begin{gather}
  \begin{split}
    &\tilde{H}_{\B} = H_{\B} + H_{\B}' = \sum_j\omega_j(\hat b^{\dagger}_j\hat b_j - \hat b^{\prime\dagger}_j\hat b_j') \equiv \sum_{k=1}^{2N}\epsilon_k\hat d_k^\dagger\hat d_k,
  \end{split}
\end{gather}
where we introduce $\epsilon_k = \omega_j$ when $\hat d_k = \hat b_j$ and $\epsilon_k = -\omega_j$ when $\hat d_k = \hat b_j'$. We further recast the environment interaction mode as 
\begin{align}
  \hat F = \sum_jc_jx_j \equiv \sum_k\zeta_k(\hat d_k + \hat d_k^\dagger),
\end{align}
with $\zeta_j = c_j\sqrt{(\bar{\mathfrak n}_j + 1)/2}$ for mode $\hat b_j$ and $\zeta_j = c_j\sqrt{\bar{\mathfrak n}_j/2}$ for mode $\hat b_j'$.

Define the ordered density operators (ODOs) as 
\begin{align}
  \rho_{\bf u,v}(t) \equiv {\rm tr}_{\B}\bigg[\mathcal N \bigg( \prod_k\hat d_k^{u_k}\hat d_k^{\dagger v_k} \bigg)\rho_{\T}(t) \bigg],
\end{align}
where $\hat d_k$ is chosen from the set $\{\hat b_j\}\cup\{\hat b_j'\}$, $\mathcal N$ is the normal ordering defined as $\mathcal N(\hat d_k^\dagger\hat d_k) = \mathcal N(\hat d_k\hat d_k^\dagger) = \hat d_k^\dagger\hat d_k$ for all modes, and the index sets are denoted as $\mathbf u \equiv \{u_k\}$ and $\mathbf v \equiv \{v_k\}$ with $u_k, v_k = 0, 1, 2,\cdots$. For later use, we also define $\mathbf u_k^\pm \equiv\{\cdots u_k \pm 1\cdots\}$ and $\mathbf v_k^\pm \equiv\{\cdots v_k \pm 1\cdots\}$. The ODOs contain the correlated and entangled properties of the system and the bath, but represent the bath degrees of freedom into a set of ordered moments. Using \Eq{vac}, we obtain the initial condition: $\rho_{\bf 0,0}(0) = \rho_{\tS}(0)$ and others are zero. Combining \Eqs{Liouville} and (\ref{bt}), we obtain the equations of motion for $\rho_{\bf u,v}$ as 
\begin{align}\label{EOM}
  \dot{\rho}_{\bf u,v} &= -i[H_{\tS}, \rho_{\bf u,v}] - i\sum_k(u_k - v_k)\epsilon_k\rho_{\bf u,v} \nl
  &\quad\, -i \sum_k\zeta_k [\hat Q, \rho_{{\bf u}_k^+,\bf v} + \rho_{{\bf u},{\bf v}_k^+}] \nl
  &\quad\, -i \sum_k\zeta_k(u_k\hat Q\rho_{{\bf u}_k^-,\bf v} - v_k\rho_{{\bf u},{\bf v}_k^-}\hat Q).
\end{align}
In deriving \Eq{EOM}, we also use the Wick's contraction concerning with the normal ordering. Using the notations, we have 
\begin{align}\label{ct}
  \la\hat F_{\B}(t)\hat F_{\B}(0)\ra_{\B} &= \sum_j \frac{c_j^2}{2}\Big[ e^{-i\omega_jt}\big( \bar{\mathfrak{n}}_j + 1 \big) + e^{i\omega_jt}\bar{\mathfrak n}_j \Big] \nl
  &\equiv \sum_k \zeta_k^2 e^{-i\epsilon_kt}.
\end{align}
We recast \Eq{ct} as 
\begin{align}\label{fdt}
  \la\hat F_{\B}(t)\hat F_{\B}(0)\ra_{\B} = \frac{1}{\pi}\int_{-\infty}^\infty\!\!\ud\omega\,e^{-i\omega t}\frac{J(\omega)}{1 - e^{-\beta\omega}},
\end{align}
where $J(\omega) = \frac{\pi}{2}\sum_jc_j^2\big[ \delta(\omega - \omega_j) - \delta(\omega + \omega_j) \big]$ is the bath spectral density, completely encapsulating information about the influence of environment, since the parameters $\{\zeta_k, \epsilon_k\}$ fully determine the dynamics of ODOs.

Equation (\ref{EOM}) is nothing but the double side hierarchical equations of motion (HEOM) for the discretized bath with correlation function given by \Eq{ct}. However, our approach differs from the original construction of HEOM in the following aspects. Firstly, we define the dynamic variables $\rho_{\bf u,v}$ without introducing the time ordering or the path integral representation of the influence functional. Secondly, the physical implements of $\rho_{\bf u,v}$ are straightforward, as the trace of one gives the corresponding ordered moments of bath modes. Thirdly and most importantly, deriving the equations of motion only utilizes the time evolution \Eq{bt} and Wick's contraction of the normal ordering, which largely overcomes the algebraic complexity of the influence functional, especially when non-linear system--bath coupling exits. (See the last part of this Letter.) It is worth noting that defining the ODOs in other ordering representations, such as anti-normal and Weyl orderings, also produce similar equations of motion as \Eq{EOM}. However, the normal ordering is the most convenient choice, since other orderings give non-zero initial conditions for $\rho_{\bf u,v}$ with $\bf u = v \neq 0$.

We turn to the quantum dissipation, where the environment becomes a thermodynamic system with continuous modes. For a continuous bath, the spectral density is assumed as a reasonably smooth function and satisfies $J(\omega\to\infty) = 0$. It usually has a simple power--law behavior \cite{Wei21,Leg871}, 
\begin{align}
  J(\omega > 0) \propto \omega^sf_{\rm c}(\omega;\omega_{\rm c})
\end{align}
and $J(-\omega) = -J(\omega)$, where $f_{\rm c}(\omega;\omega_{\rm c})$ is a cutoff function with $\omega_{\rm c}$ being the cutoff frequency. One straightforward way to the dissipative dynamics is to evolute \Eq{EOM} with $N$ being a large number. The parameters $\{\zeta_k, \epsilon_k\}$ are obtained by discretizing the spectral density. However, this approach is computationally expensive, and thus nonrealistic and impractical. The other approach utilizes the fluctuation--dissipation theorem [\Eq{fdt}] to expand the bath correlation function in terms of a series of exponential functions, 
\begin{equation}\label{fexp}
  \begin{split}
    \la\hat F_{\B}(t)\hat F_{\B}(0)\ra_{\B} &\simeq \sum_{k=1}^K\eta_ke^{-\gamma_kt},\\
    \la\hat F_{\B}(0)\hat F_{\B}(t)\ra_{\B} &\simeq \sum_{k=1}^K\eta_{k}^*e^{-\gamma_k^*t},
  \end{split}
\end{equation}
with $t>0$ and $\{\eta_k, \gamma_k\}$ being complex. 
Equation (\ref{fexp}) becomes exact when $K$ goes to infinity. For the cutoff function being a rational function, the exponential decomposition is evaluated via the Cauchy's residue theorem in contour integration. The integration via residues depends on not only the concrete form of the spectral density, but also the fractional decomposition of the bosonic function. For the latter, traditionally, people adopt the Mittag--Leffler decomposition, specifically named also as the Matsubara expansion.
Besides, the Pad\'{e} spectrum decomposition (PSD) \cite{Hu10101106,Hu11244106} can greatly decrease the number of decomposition terms of the bosonic function part for the same precision. By far, one of the most efficient and powerful expansions of \Eq{fexp} is the time-domain Prony fitting decomposition ($t$-PFD) \cite{Che22221102}, which fits the time correlation function with the minimum terms and is applied to arbitrary spectral density functions---including rational functions, exponential functions, step functions, etc. 

The decomposed bath correlation function \Eq{fexp} presents a thermofield quasi-particle picture of the bath influence, satisfying 
\begin{gather}
  \hat F = \sum_k(\sqrt{\eta_k} \hat f_k + \sqrt{\eta_k^*}\hat f_{k}^\dagger),\quad \hat f_k|{\rm Vac}\ra = 0,
\end{gather} 
and time evolution,
\begin{align}\label{gdiff}
  \hat f_k^{\B}(t) \equiv e^{i\tilde{H}_{\B}t}\hat f_ke^{-i\tilde{H}_{\B}t} = \hat f_ke^{-\gamma_kt}.
\end{align}
And each $\hat f_k$ is a bosonic operator independent of the others, that is $[\hat f_k, \hat f_{k'}] = [\hat f_k, \hat f_{k'}^\dagger] = 0$ for $k\neq k'$. However, the Wick's contraction of the normal ordering is not as simple as the discrete bath case, since the exponential decomposition with complex parameters violates the time translational symmetry, i.e., $\la\hat F_{\B}(t)\hat F_{\B}(0)\ra_{\B} \neq \la\hat F_{\B}(0)\hat F_{\B}(-t)\ra_{\B}$. Then from \Eq{fexp}, we have 
\begin{align}\label{gWick}
  \begin{split}
    &{\rm tr}_{\B}\big(\hat f_k^>\hat f_{k'}^{\dagger>}\rho_{\B}^{\rm eq}\big) = e^{i\theta_k}\delta_{kk'} + {\rm tr}_{\B}\big[\mathcal N(\hat f_k\hat f_{k'}^{\dagger})\rho_{\B}^{\rm eq}\big], \\ 
    &{\rm tr}_{\B}\big(\hat f_k^{\dagger>}\hat f_{k'}^{<}\rho_{\B}^{\rm eq}\big) = e^{-i\theta_k}\delta_{kk'} + {\rm tr}_{\B}\big[\mathcal N(\hat f_{k'}\hat f_{k}^{\dagger})\rho_{\B}^{\rm eq}\big].
  \end{split}
\end{align}
Here, $\theta_k\equiv \arg\eta_k$ is the phase of $\eta_k$, $\hat f_k^\gtrless$ are superoperators in the Liouville space, defined as $\hat f_k^{>}\hat O \equiv \hat f_k\hat O$, $\hat f_k^<\hat O\equiv \hat O\hat f_k$, and the same for $\hat f_{k}^{\dagger\gtrless}$. When $\{\eta_k\}$ are real numbers, \Eq{gWick} reduces to the original Wick's contraction. \Eq{gWick} is thus treated as a generalization of the original Wick's contraction. This is because non-unitary evolutions with time arrow can be only described in the Liouville space instead of the Hilbert space \cite{Geo73277}.

Define the ordered density operators in the $\{\hat f_k\}$-representation as
\begin{align}
  \rho_{\bf u,v}(t) \equiv {\rm tr}_{\B}\bigg[\mathcal N \bigg( \prod_{k=1}^K\hat f_k^{u_k}\hat f_k^{\dagger v_k} \bigg)\rho_{\T}(t) \bigg].
\end{align}
Using \Eqs{Liouville}, (\ref{gdiff}), and (\ref{gWick}), we readily derive the equations of motion, reading
\begin{align}\label{HEOM}
  \dot{\rho}_{\bf u,v} &= -i[H_{\tS}, \rho_{\bf u,v}] - \sum_k(u_k\gamma_k + v_k\gamma_k^*)\rho_{\bf u,v} \nl
  &\quad\, -i \sum_k[\hat Q, \sqrt{\eta_k} \rho_{{\bf u}_k^+,\bf v} + \sqrt{\eta_k^*} \rho_{{\bf u},{\bf v}_k^+}] \nl
  &\quad\, -i \sum_k(u_k\sqrt{\eta_k}\hat Q\rho_{{\bf u}_k^-,\bf v} - v_k\sqrt{\eta_k^*}\rho_{{\bf u},{\bf v}_k^-}\hat Q),
\end{align}
which is exactly the double side HEOM \cite{Xu22230601}. The initial state is given by, $\rho_{\bf 0,0}(0) = \rho_{\tS}(0)\delta_{\bf u,0}\delta_{\bf v,0}$. Notice that most of the exponential decomposition strategies satisfy the pairing condition: the complex conjugation $\gamma_k^* \equiv \gamma_{\bar k}$ also belongs to the exponent set $\{\gamma_k\}$. One may simplify \Eq{HEOM} into the single side HEOM by defining
\begin{align}\label{ddo}
  \rho_{\bf n}(t) \equiv {\rm tr}_{\B}\bigg[ \mathcal N\bigg( \prod_{k=1}^K\hat\phi_k^{n_k} \bigg)\rho_{\T}(t) \bigg]
\end{align}
with $\hat\phi_k \equiv \sqrt{\eta_k}\hat f_k + \sqrt{\eta_{\bar k}^*}\hat f_{\bar k}^\dagger$ and ${\bf n} \equiv \{n_k| n_k = 0,1,2,\cdots\}$. Then we have 
\begin{align}
  \dot\rho_{\bf n} &= -i[H_{\tS}, \rho_{\bf n}] - \sum_kn_k\gamma_k\rho_{\bf n} -i \sum_k[\hat Q,\rho_{{\bf n}_k^+}] \nl
  &\quad\, -i \sum_k n_k\big(\eta_k\hat Q\rho_{{\bf n}_k^-} - \eta_{\bar k}^*\rho_{{\bf n}_{\bar k}^-}\hat Q\big).
\end{align}
Here, we have used $\hat\phi_k^{\B}(t) = e^{i\tilde H_{\B}t}\hat\phi_ke^{-i\tilde{H}_{\B}t} = \hat\phi_k e^{-\gamma_kt}$. In the previous work, the quasi-particles described by $\{\hat\phi_k\}$ are named as \textit{dissipatons} \cite{Yan14054105,Wan22170901}, etymologically derived from the verb ``dissipate'' and the suffix ``-on''. The dynamic variables $\rho_{\bf n}$ are also named as the dissipaton density operators. We can evaluate the system--bath correlated dynamics by utilizing the generalized Wick's contraction \Eq{gWick}, e.g., ${\rm Tr}(\hat A_{\tS}\hat F^2\rho_{\T}) = \sum_k\eta_k{\rm tr}_{\tS}(\hat A_{\tS}\rho_{\tS}) + \sum_{kk'}{\rm tr}_{\tS}(\hat A_{\tS}\rho_{{\bf 0}_{kk'}^{++}})$ for any system operator $\hat A_{\tS}$. Furthermore, the definition of ODOs provide a holographic mapping of the original Liouville space to the linear space expanded by $\rho_{\bf n}$. See Ref.\,\cite{Wan22170901} for more details. The dissipatons also play roles as generalized Brownian particles, with the collective dynamics of the system and disspatons being a generalized Zusman master equation form \cite{Li23214110,Li24032620}. 

To end this Letter, we finally show our theory's ability to handle nonlinear system--bath interactions---beyond the Gauss--Wick condition. 
Consider the total Hamiltonian with interaction being $H_{\SB} = \alpha_0\hat Q_0 + \alpha_1\hat Q_1\hat F + \alpha_2\hat Q_2\hat F^2$. Define the dynamic variables still as \Eq{ddo}. By applying the Wick's contraction twice, we obtain the extended equations of motion for quadratic coupling \cite{Che23074102},
\begin{align}
  \dot\rho_{\bf n} &= -i[H_{\tS} + \alpha_0\hat Q_0 + \alpha_2\la\hat F^2\ra_{\B}\hat Q_2, \rho_{\bf n}] - \sum_kn_k\gamma_k\rho_{\bf n} \nl
  &\quad\, -i \alpha_1\sum_k[\hat Q_1,\rho_{{\bf n}_k^+}] -i \alpha_2\sum_{kk'}[\hat Q_2,\rho_{{\bf n}_{kk'}^{++}}] \nl
  &\quad\, -i\alpha_1 \sum_k n_k\big(\eta_k\hat Q_1\rho_{{\bf n}_k^-} - \eta_{\bar k}^*\rho_{{\bf n}_k^-}\hat Q_1\big) \nl
  &\quad\, -i\alpha_2 \sum_{kk'}n_k(n_{k'} - \delta_{kk'})\nl
  &\qquad\qquad\times\big( \eta_k\eta_{k'}\hat Q_2\rho_{{\bf n}_{kk'}^{--}} - \eta_{\bar k}^*\eta_{\bar k'}^*\rho_{{\bf n}_{kk'}^{--}}\hat Q_2 \big).
\end{align}
Extending to higher order couplings is straightforward.

\vspace{1em}

Support from the National Natural Science Foundation of China (Grant Nos.\,224B2305) is gratefully acknowledged.


%

\end{document}